\documentclass{article}
\usepackage{ijcai22}

\usepackage{times}
\usepackage{soul}
\usepackage{url}
\usepackage[hidelinks]{hyperref}
\usepackage[utf8]{inputenc}
\usepackage[small]{caption}
\usepackage{graphicx}
\usepackage{amsmath,amssymb}
\usepackage{amsthm}
\usepackage{comment}
\usepackage{booktabs}
\usepackage{algorithm}
\usepackage{algorithmic}
\usepackage{enumitem}
\usepackage[mathscr]{euscript}
\usepackage{xcolor}

\newcommand{\swapna}[1]{\textcolor{magenta}{(#1 --Swapna)}}
\newcommand{\hsm}[1]{\textcolor{red}{#1 (HSM)}}
\newcommand{\mvm}[1]{\textcolor{green}{(#1 (Madhav)}}

\newtheorem{proposition}{Proposition}

\title{A  Reliability-aware Distributed Framework to Schedule Residential Charging of Electric Vehicles}
\author{
Rounak Meyur$^1$\footnote{Contact Author}\and
Swapna Thorve$^1$\and
Madhav Marathe$^1$\and
Anil Vullikanti$^1$\and
Samarth Swarup$^1$\and
Henning Mortveit$^1$\\
\affiliations
$^1$UVA Biocomplexity Institute Charlottesville, Virginia, USA\\
\emails
\{rm5nz, st6ua, marathe, vsakumar, swarup, henning.mortveit\}@virginia.edu,
}

\begin{document}
\maketitle

\begin{abstract}
%With a goal of zero carbon emission, the modern day electric power grid is experiencing rapid advancements through the deployment of distributed energy resources (DERs) and electric vehicles (EVs). 
Residential consumers have become active participants in the power distribution network after being equipped with residential EV charging provisions. This creates a challenge for the network operator tasked with dispatching electric power to the residential consumers through the existing distribution network infrastructure in a reliable manner. 
In this paper, we address the problem of scheduling residential EV charging for multiple consumers while maintaining network reliability. An additional challenge is the restricted exchange of information: where the consumers do not have access to network information and the network operator does not have access  to consumer load parameters. 
We propose a distributed framework which generates an optimal EV charging schedule for individual residential consumers based on their preferences and iteratively updates it until the network reliability constraints set by the operator are satisfied. 
We  validate the proposed approach for different EV adoption levels in a synthetically created digital twin of an actual power distribution network. The results demonstrate that the new approach can achieve a higher level of network reliability compared to the case where residential consumers charge EVs based solely on their individual preferences, thus providing a solution for the existing grid to keep up with increased adoption rates without significant investments in increasing grid capacity.
\end{abstract}

\section{Introduction}\label{sec:intro}

Studies have shown that home-charging units are pivotal infrastructure for promoting EV adoption~\cite{Wei2021}. 
As EV adoption increases over the next few years, the power drawn from the grid will increase and may cause disturbances in the distribution power network.
It is quite possible that the network operation may undergo modifications in order to accommodate these unconventional loads in the future without affecting the reliability of the grid.

From the standpoint of power engineering, a \emph{reliable} power grid is one which has adequate generation to support the consumer load demand and can be operated without violating standard power engineering constraints~\cite{Billinton1994}. 
In this paper, we consider an operational problem and therefore are not concerned about the aspect of adequate generation.
%(which is usually dealt in planning problems). 
Thus, we use the term \emph{reliable distribution network} to mean a network that can satisfy the loads without violating the node voltages and line flow (edge flow) capacities.
%\swapna{Rounak, I think you should rename edge flows in experiments and results to line flow capacities. Make sure to add those change sto the plots as well.}
%to be representative of a distribution grid being operated with node voltages and line power flows within the acceptable limits. 
While flows above line rating causes overheating of conductors and subsequent physical damage, node voltages represent the quality of power delivered at the node. Undervoltage and overvoltage at residence nodes lead to eventual failure of household appliances. Relevant basic concepts of power distribution networks are described in the Appendix.

Traditionally, the distribution network is designed to sustain the peak load demand of consumers~\cite{dist_plan}. The predictable growths in consumer peak demands and energy consumption enables the network operator to plan and operate the network reliably. However, adoption of EVs in residential communities leads to a significant deviation in the predictability of consumer loads~\cite{ev-problem1}. 
Residential EV charging constitutes a significant percentage of net household demand leading to large power consumed from the grid.
The problem is more dominant when residential consumers opt to charge EVs according to their personal convenience~\cite{ev-problem2}. 
The excess load consumed by the EV charging units adversely affects distribution grid by causing transformer overloading or high voltage drop at feeder ends~\cite{ev-problem3}. 
As a result, it is desirable to develop a framework which aids residential consumers with their goal of scheduling EV charging based on their individual preferences, and simultaneously taking into account the grid reliability requirements of the network operator.

\noindent \textbf{Contributions.}~The contributions of the paper are summarized here: (i) A novel `reliability-aware distributed EV charging scheduling framework' is proposed. It uses information such as the hourly electricity rate, household energy demand profiles \& their preferences as inputs for consumers, and power engineering constraints of distribution network as inputs for operator to aid residential EV adopters in scheduling their EV charging units in an optimal manner without affecting the reliability of the power grid. (ii) The distributed framework uses ADMM based iterative methodology which guarantees an optimal solution for our problem. Each iteration involves solving a mixed integer quadratic program (MIQP) for each residential consumer and a quadratic program (QP) for the operator. The optimal solutions are exchanged and used in succeeding iterations until a consensus is reached. This minimum exchange of information between the consumers/residences/households and the network operator can be executed using present smart grid infrastructure and avoids sharing of private and proprietary data. (iii) We use digital duplicates of residential consumer load demand profile and power distribution networks resembling the actual physical counterparts for our case studies. This facilitates conducting real-world test scenarios to explore the impact of using the proposed framework while considering multiple levels of EV adoption. Our experiment results demonstrate that the proposed distributed framework helps maintain  network reliability compared to the case where EV adopters charge their vehicles based on their personal (individualized) preferences.

\begin{comment}
\begin{enumerate}
%framework
    \item A novel `distributed EV charging scheduling framework for the power distribution network' is proposed. It uses information such as the hourly electricity rate, household energy demand profiles \& their preferences, power engineering constraints of distribution networks, and fixed time-slots available for EV charging to enable residential EV adopters to schedule their EV charging units in an optimal manner without affecting the reliability of the power grid.
% method in the framework
    \item The distributed framework uses ADMM based iterative methodology which guarantees an optimal solution. Each iteration involves solving a mixed integer quadratic program (MIQP) for each residential consumer and a quadratic program (QP) for the operator. The optimal solutions are exchanged and used in succeeding iteration until a consensus is reached. This minimum exchange of information between the consumers/residences/households and the network operator can be executed using present smart grid infrastructure. 
% realistic scenario simulations
    \item We use digital duplicates of residential consumer load demand profile and digital duplicate of power distribution networks resembling the actual physical counterparts.
    This facilitates conducting real-world test scenarios to explore the impact of using the proposed framework while considering multiple levels of EV adoption.
\end{enumerate}
\end{comment}

\section{Related Work}\label{sec:relate}

Several works have been presented in the literature for scheduling EV charging. In general, optimization techniques are a popular choice for solving the problem of scheduling EV charging at household level~\cite{Cao2012,Zhao2018,Lee2020,Blonsky2021,Young-Min2013,Wanrong2016,goncalves2018,Khonji2018,Nyns2010,sairaj2014}. 
%7383228,Gan2012,Kikusato2019
%Gao_2021,Lee2019
Recently, ML techniques such as neural networks~\cite{Shuvo2021} and reinforcement learning frameworks~\cite{Yongsheng2022} have been proposed to study the problem of scheduling EV charging loads in smart homes.

%Khonji et al. develop an approximation algorithm and a fast heuristic to solve the scheduling optimization problem of EV charging~\cite{Khonji2018}.
%A genetic algorithm optimization framework is used to schedule different types of loads in a household such as HVAC, appliances, energy storage system, and EV~\cite{goncalves2018}.
%Stochastic optimization techniques such as quadratic programming and dynamic programming are proposed for coordinated EV charging with the goal of minimizing the power losses and to maximize the main grid load factor~\cite{Nyns2010}.
%Gan et al.~\cite{Gan2012} formulate EV charging scheduling as a discrete optimization problem and is solved in an iterative fashion via communication between EV and transformer.

Most of these works implement a centralized approach to schedule EV charging.
A centralized optimization algorithm/framework evaluates the optimal power consumption patterns which are beneficial to only one of the entities -- consumers or network operators~\cite{Liu2015}. 
This approach may not be realistic since details of the individual consumer load is usually not  accessible to the network operator. 
At the same time, the network topology and parameters are unknown to the consumers. 
Under such circumstances, a de-centralized/distributed framework is useful since it can help network operators and consumers communicate essential information that can respect both -- network reliability and consumer preferences. 
The current smart grid infrastructure supports the development of such a framework due to the availability of two-way communication.

Dall’Anese \emph{et al.}~\shortcite{sairaj2014} use an alternating direction method of multipliers (ADMM) based approach to evaluate inverter set points at different locations in a network while maintaining network reliability. The results show that this method provides superior convergence guarantees in comparison with other methods while dealing with mixed integer linear programs (MILP). In this work, we propose a distributed optimization framework based on the ADMM method for scheduling EV charging in a power distribution network. Our framework satisfies two goals -- maintaining grid reliability for the network operator while respecting consumer preferences.

\section{Problem Formulation}\label{sec:problem}
In this paper, we are interested in power consumption trajectory over a finite horizon time window of $T$ intervals from time instant $k=0$ to $k=T$. We define an interval $t$ as the duration between time instants $k=t-1$ and $k=t$. Table~\ref{tab:set-index} summarizes the index variables and sets used in the paper.
\begin{table}[htbp]
\centering
\caption{Summary of index variables and sets used}
\label{tab:set-index}
\begin{tabular}{ll}
\hline
\textbf{Symbol} & \multicolumn{1}{c}{\textbf{Description}} \\ \hline
$T$ & Number of intervals in time window \\
$N$ & Number of non-substation nodes in network \\
$i$ & Index of node in network \\
$t$ & Index of time interval \\
$k$ & Index of time instant \\
$\alpha,\beta$ & Limits of squared voltage magnitude \\
$\mathscr{V}$ & Set of all nodes in network \\
%$\mathscr{E}$ & Set of all edges in network \\
$\mathscr{H}$ & Set of all residence nodes in network \\
$\mathscr{N}$ & Set of all non-substation nodes in network \\\hline
\end{tabular}
\end{table}
\subsection{Distribution Network Model}
The power distribution network is a \emph{tree} comprising of $N+1$ nodes collected in the set $\mathscr{V}:=\mathscr{N}\cup\{0\},\mathscr{N}:=\{1,2,\cdots,N\}$. The tree is rooted at substation node $\{0\}$ and consists of primary and secondary distribution lines collected in the edge set. The set $\mathscr{N}$ includes residences, local transformers and auxiliary nodes required to connect the transformers and residences~\cite{rounak2020,kersting_book,Bolognani2015}. Here, we are interested in the variables associated with the set of residence nodes $\mathscr{H}\subset\mathscr{N}$. The power consumption and squared voltage magnitude at node $i$ and time $t$ are denoted respectively by $p_{i}^{t}$ and $v_{i}^{t}$. We respectively stack these variables for $i\in\mathscr{N}$ to corresponding vectors $\mathbf{p}^t$ and $\mathbf{v}^t$ for every time interval $t$. The non-linear relation between power injections and squared voltages in the network can be simplified to linear expression using the Linearized Distribution Flow (LDF) model~\cite{Bolognani2015}.
\begin{subequations}
\begin{align}
    & \mathbf{v}^t = -2\mathbf{Rp}^t + \mathbf{1} \label{seq:volt-power}\\
    & \alpha \mathbf{1} \leq \mathbf{v}^t \leq \beta \mathbf{1} \label{seq:volt-limit}
\end{align}
\label{eq:volt-constraint}
\end{subequations}
Here, $\mathbf{1}$ is a vector of all $1$'s. The matrix $\mathbf{R}$ is a function of network topology and edge parameters which is only accessible by the network operator. The primary objective of the operator is to maintain the network reliability where the node voltages are within acceptable ANSI C.84 Range A limits~\cite{ansi}. In most practical distribution networks, these limits are 0.95 pu and 1.05 pu. The operator ensures that the power consumption at different nodes in the network satisfy (\ref{eq:volt-constraint}) where $\alpha,\beta$ denote the squared voltage limits.

\subsection{Residence Load Models}
This section describes load demand of a residence node $i\in\mathscr{H}$. The aggregate power consumption for the time interval $t$ is given by $p_{i}^t\in\mathbb{R}$. This load comprises of an uncontrollable base load demand denoted by $p_{i,0}^{t}$ and the controllable counterpart. In this paper, we use the synthetically generated residential load demand data described in~\cite{swapna2018} for the uncontrollable base load demand. We consider residence owned EV charging stations as the only controllable load which is denoted by $p_{i,\textrm{EV}}^{t}$. The power consumption at node $i$ and over time interval $t$ can be expressed as
\begin{equation}
    p_{i}^{t} = p_{i,0}^{t} + p_{i,\textrm{EV}}^{t} \quad \forall t=1,2,\cdots,T \label{eq:load-bal}
\end{equation}

\noindent\textbf{EV Charging Model.}~We assume that the EV charging unit is responsible to charge only a single EV owned by the customer. Let the charge capacity of the EV be $Q_{i,\textrm{EV}}$ and the power rating of the EV charging unit be $P_{i,\textrm{EV}}$. The state of charge (SOC) evolves over the time interval $t$ from $s_{i,\textrm{EV}}^{t-1}$ to $s_{i,\textrm{EV}}^{t}$ following (\ref{seq:ev-evol}). Further, the constraint (\ref{seq:ev-lim}) limits the SOC to suitable lower and upper bounds. The scheduling problem aims to find out the optimal time intervals when the EV can be charged while maintaining a secure power grid. Let $z_{i,\textrm{EV}}^t\in\{0,1\}$ be a binary variable which takes the value $1$ if the EV is charged at time interval $t$ and $0$ otherwise. 

Further, the EV is available for charging only at particular time interval denoted by the closed interval $\mathscr{T}_{i,\textrm{EV}}:=\left[t_{\textrm{start}},t_{\textrm{end}}\right]$. Let the SOC at time $t=t_{\textrm{start}}$ be $s_{i,\textrm{init}}$ and it is expected that by the end of the interval, the SOC of the battery needs to be at least $s_{i,\textrm{final}}$.  Note that we consider a very simple case where the EV is available to be charged within a single continuous time interval $\mathscr{T}_{i,\textrm{EV}}$ in the time window of $T$ intervals. We also assume that the EV is not used in this interval. In realistic scenarios, these intervals of charging are discontinuous and usage of EV would result in different SOC at different time intervals.
\begin{subequations}
\begin{align}
    & p_{i,\textrm{EV}}^{t} = z_{i,\textrm{EV}}^tP_{i,\textrm{EV}} & \forall t=1,2,\cdots,T \label{seq:ev-power}\\
    & s_{i,\textrm{EV}}^{t} = s_{i,\textrm{EV}}^{t-1} + \frac{p_{i,\textrm{EV}}^{t}}{Q_{i,\textrm{EV}}} & \forall t=1,2,\cdots,T \label{seq:ev-evol}\\
    & 0 \leq s_{i,\textrm{EV}}^{k} \leq 1 & \forall k=0,1,\cdots,T \label{seq:ev-lim}\\
    & z_{i,\textrm{EV}}^{t} = 0 & \forall t \notin \mathscr{T}_{i,\textrm{EV}} \label{seq:ev-nocharge}\\
    & s_{i,\textrm{EV}}^{t_\textrm{start}} = s_{i,\textrm{init}},\quad s_{i,\textrm{EV}}^{t_\textrm{end}} \geq s_{i,\textrm{final}} & \label{seq:ev-init}
    %& s_{i,\textrm{EV}}^{t_\textrm{end}} \geq s_{i,\textrm{final}} & \label{seq:ev-final}
\end{align}
\label{eq:res-ev}
\end{subequations}

\subsection{Optimization problem}
Each residence aims to compute the optimal power usage trajectory of its EV charging unit over a finite horizon time window of length $T$ denoted by $\{p_{i,\textrm{EV}}^{t}\}_{t=1}^{T}$. Given the hourly rate of electricity $c^{t}$ for each time interval in the time window, the optimization problem for each residence involves minimizing the total cost of consumption given the EV constraints (\ref{eq:res-ev}). This results in the MILP (\ref{eq:ind-opt}).
\begin{subequations}
\begin{align}
    \min \quad & \sum_{t=1}^{T} c^t p_{i}^{t} \\
    \textrm{over} \quad & p_{i,\textrm{EV}}^{t}  \quad\quad\quad\quad~~~ \forall t\\
    \textrm{s.t.} \quad & (\ref{eq:load-bal}),(\ref{eq:res-ev})  ~~~\quad\quad\quad \forall t
\end{align}
\label{eq:ind-opt}
\end{subequations}
At the same time, the network operator needs to ensure node voltages are within acceptable limits to maintain a reliable system. Additionally the operator might aim to optimize other aspects such as minimize losses or reduce voltage deviation~\cite{sairaj2014}. This can be expressed by $C\left(\mathbf{p}^t\right)$ which is a function of power usage of all residences at interval $t$. In this paper, we do not consider any particular objective of the network operator and treat $C\left(\mathbf{p}^t\right)=0$. We define the \emph{Reliability-aware EV Charge Scheduling}({\sc REVS}) problem which satisfies consumer preferences as well as ensures network reliability by (\ref{eq:total-opt}).
\begin{subequations}
\begin{align}
    \min \quad & \sum_{t=1}^{T}C\left(\mathbf{p}^t\right) + \sum_{i\in\mathscr{H}}\sum_{t=1}^{T} c^t p_{i}^{t} \\
    \textrm{over} \quad & p_{i,\textrm{EV}}^{t}  \quad\quad\quad\quad~~~ \forall t~ \forall i\in\mathscr{H}\\
    \textrm{s.t.} \quad & (\ref{eq:volt-constraint}) \quad\quad\quad\quad\quad~~ \forall t\\
    & (\ref{eq:load-bal}),(\ref{eq:res-ev})  ~~~\quad\quad\quad \forall t~~ \forall i\in\mathscr{H}\\
    & p_{i}^{t} = 0 \quad\quad\quad\quad \forall t~~ \forall i\notin\mathscr{H}
\end{align}
\label{eq:total-opt}
\end{subequations}

%\anil{$C(p)$ is a bit confusing above. If you treat $C(p)=0$, can't you drop it from the objective?}
%\rounak{Yes we can drop it. But this is the generic form and I wanted to keep it.}
\section{Proposed Methodology}
The {\sc REVS} problem in (\ref{eq:total-opt}) is an MILP with binary variables arising from the on/off status of the EV charging unit if we consider . This problem can be solved from a central location (such as the operator) if the load information of residences are known. However, this is not always the case due to privacy concern associated with sharing personal data of consumers. Similarly, the network topology and parameters are considered proprietary information and cannot be shared with the consumers.
However, limited information exchange such as total power consumption can be done without violating privacy concerns using the current smart grid infrastructure. In this section, we propose an iterative method based on the ADMM technique to reach the optimal solution for the {\sc REVS} problem.
%\anil{might be useful to motivate better}

To this end, we separate the problem for the network operator and individual residences. Each residence $i$ aims to compute the optimal power usage trajectory $\left\{p_{i}^t\right\}_{t=1}^T$ over the time window given the EV charging constraints. The network operator computes consumption trajectories $\left\{\tilde{p}_i^t\right\}$ for all nodes (in the vector form $\tilde{\mathbf{p}}^t$) such that the network reliability constraints are satisfied. Additionally, we add constraint (\ref{seq:admm-equal}) to force these trajectories to match each other. Therefore, we get (\ref{eq:total-opt-alternate}) as the alternate version of the {\sc REVS} problem. 
%Note that $\tilde{\mathbf{p}}^t,\mathbf{p}^t$ respectively denote the vectors of all $\tilde{p}_i^t$ and $p_i^t$ of all nodes in the network stacked together for a particular time interval $t$.
\begin{subequations}
\begin{align}
    \min \quad & \sum_{i\in\mathscr{H}}\sum_{t=1}^{T} c^t p_{i}^{t} \\
    \textrm{over} \quad & p_{i}^{t},\tilde{p}_{i}^{t}  ~~\quad\quad\quad\quad~ \forall t~~ \forall i\\
    \textrm{s.t.} \quad & (\ref{eq:volt-constraint}) ~\quad\quad\quad\quad\quad~~ \forall t\\
    & (\ref{eq:load-bal}),(\ref{eq:res-ev})  \quad\quad\quad\quad \forall t~~ \forall i\in\mathscr{H}\\
    & p_{i}^{t} = 0 = \tilde{p}_i^t ~\quad\quad \forall t~~ \forall i\notin\mathscr{H}\\
    & \tilde{p}_i^t = p_{i}^t ~\quad\quad\quad~~~ \forall t ~~ \forall i\in\mathscr{H}\label{seq:admm-equal}
\end{align}
\label{eq:total-opt-alternate}
\end{subequations}
\begin{table}[htbp]
\centering
\caption{Summary of variables in optimization problem}
\label{tab:opt-var}
\begin{tabular}{ll}
\hline
\textbf{Symbol} & {\textbf{Description}} \\ \hline
$v_i^t$ & Voltage at node $i$ for interval $t$ \\
$p_i^t$ & Power consumed at node $i$ for interval $t$\\
$\tilde{p}_i^t$ & Power consumption computed by operator \\
$p_{i,0}^t$ & Power consumed by fixed load \\
$p_{i,\textrm{EV}}^t$ & Power consumed by EV charging unit \\
$z_{i,\textrm{EV}}^t$ & On/Off status of EV charging unit \\
$s_{i,\textrm{EV}}^k$ & SOC of EV at time instant $k$ \\
$\gamma_i^t$ & Dual variable corresponding to (\ref{seq:admm-equal})\\
$\mathbf{v}^t$ & Vector of $v_i^t$ for all nodes $i\in\mathscr{N}$ \\
$\mathbf{p}^t$ & Vector of $p_i^t$ for all nodes $i\in\mathscr{N}$ \\
$\tilde{\mathbf{p}}^t$ & Vector of $\tilde{p}_i^t$ for all nodes $i\in\mathscr{N}$\\
%$\tilde{\mathbf{p}}$ & Collection of $\tilde{\mathbf{p}}^t$ for all intervals $t$\\
\hline
\end{tabular}
\end{table}

\begin{figure}[!ht]
    \centering
    \includegraphics[width=0.48\textwidth]{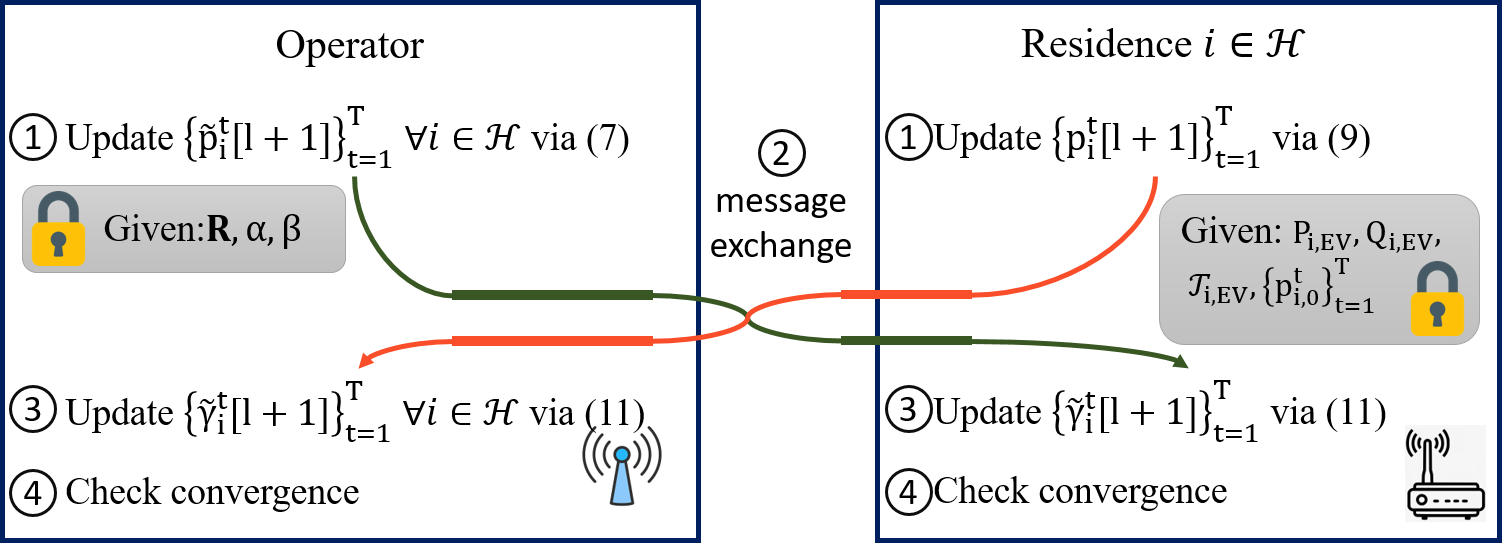}
    \caption{Message exchange aided proposed distributed framework to schedule residential EV charging.}
    \label{fig:message-xchange}
\end{figure}

Now, we use the conventional ADMM steps to iteratively update the optimization variables for the operator and residences. Let $\tilde{\mathcal{P}}[l]:=\left\{\tilde{\mathbf{p}}^t[l]\right\}_{t=1}^{T}$ denote the optimal trajectories for all nodes computed by the operator for iteration $l$. Similarly, let $\mathcal{P}_i[l] := \left\{{p}_i^t[l]\right\}_{t=1}^{T}$ denote the optimal power usage trajectory computed by residence $i$. We abuse the notation $\left\{\mathcal{P}_i[l]\right\}$ to denote the optimal trajectories $\left\{{p}_i^t[l]\right\}_{t=1}^{T}$ computed by all residences $i\in\mathscr{H}$ individually. The two steps of iteration are listed below. Note that the first step is carried out simultaneously for all residences and network operator. Fig.~\ref{fig:message-xchange} illustrates the proposed message passing based distributed framework.
%iteratively solve the following problems and update the dual variables $\gamma_i^t$. Note that the dual variables correspond to the equality constraint (\ref{seq:admm-equal}) for the Lagrangian function of the optimization problem in (\ref{eq:total-opt-alternate}). We consider iteration index $l+1$ where we use the results of iteration $l$ to compute the power trajectories $\tilde{\mathbf{p}}[l+1]$ and $\mathbf{p}[l+1]$. 
\begin{enumerate}[font=\bfseries]
    \item[S1a.] At the operator side, we update the operator estimated power consumption $\tilde{p}_i^t$ for all residences using (\ref{eq:opt-operator}).
    \begin{subequations}
    \begin{align}
        \tilde{\mathcal{P}}[l+1] := &\textrm{arg}~\textrm{min}  \quad F(\tilde{\mathcal{P}}[l],\{\mathcal{P}_i[l]\}) \label{seq:obj-operator}\\
        &\textrm{s.to.} \quad \alpha \leq 1-2\sum_{j=1}^NR_{ij}\tilde{p}_{j}^{t} \leq \beta \quad \forall t~~\forall i\label{seq:volt-operator}
    \end{align}
    \label{eq:opt-operator}
    \end{subequations}
    where the function $F(\tilde{\mathcal{P}}[l],\{\mathcal{P}_i[l]\})$ is defined as
    \begin{equation}
    \begin{aligned}
        F(\tilde{\mathcal{P}}[l],&\{\mathcal{P}_i[l]\}) := \sum_{i\in\mathscr{H}}\sum_{t=1}^{T}\frac{\kappa}{2}\left(\tilde{p}_{i}^t\right)^2 \\
        + & \sum_{i\in\mathscr{H}}\sum_{t=1}^{T}\tilde{p}_{i}^t \left(\gamma_{i}^{t}[l] - \frac{\kappa}{2}\tilde{p}_{i}^t[l]-\frac{\kappa}{2}p_{i}^{t}[l]\right)
    \end{aligned}
    \end{equation}
    \item[S1b.]For residence $i$, we update using (\ref{eq:opt-residence}).
    \begin{subequations}
    \begin{align}
        \mathcal{P}_{i}[l+1] := \textrm{arg}~\textrm{min} \quad & \sum_{t=1}^{T} c^t p_{i}^{t} + F_{i}(\tilde{p}_{i}^{t}[l],p_{i}^{t}[l])& \label{seq:obj-residence}\\
        \textrm{s.to.} \quad & (\ref{eq:load-bal})-(\ref{eq:res-ev}) &
    \end{align}
    \label{eq:opt-residence}
    \end{subequations}
    where the function $F_i\left(\tilde{p}_{i}^{t}[l],p_{i}^{t}[l]\right)$ is defined as
    \begin{equation}
    \begin{aligned}
        F_{i}\left(\tilde{p}_{i}^{t}[l],p_{i}^{t}[l]\right) = & \sum_{t=1}^{T}\frac{\kappa}{2}\left({p}_{i}^t\right)^2 \\ - & \sum_{t=1}^{T}{p}_{i}^t \left(\gamma_{i}^{t}[l] + \frac{\kappa}{2}\tilde{p}_{i}^t[l]+\frac{\kappa}{2}p_{i}^{t}[l]\right)
    \end{aligned}
    \label{eq:f_res}
    \end{equation}
    \item[S2.]At the operator and residence sides, the dual variable update is updated.
    \begin{equation}
        \gamma_{i}^{t}[l+1] = \gamma_{i}^{t}[l] + \frac{\kappa}{2}\left(\tilde{p}_{i}^{t}[l+1]-p_{i}^{t}[l+1]\right)
        \label{eq:gamma}
    \end{equation}
\end{enumerate}
The resulting decentralized procedure involves a two-way message exchange of the iterates $\left\{\tilde{\mathbf{p}}^t[l]\right\}_{t=1}^{T}$ and $\left\{\mathbf{p}^t[l]\right\}_{t=1}^{T}$ between the network operator and residential consumers. At an iteration $l>0$, the network operator updates the power trajectories based on (\ref{eq:opt-operator}) whose objective includes a regularization term $F(\tilde{\mathcal{P}}[l],\{\mathcal{P}_i[l]\})$. This term enforces consensus with the power usage trajectories computed at the residences. The constraints ensure the reliability aspects in of the network. Note that (\ref{eq:opt-operator}) is a QP because of the quadratic regularization term. The operator relays to each residential consumer $i$ a copy of the iterate value $\left\{\tilde{p}_i^t[l+1]\right\}_{t=1}^{T}$. 
At the same time, the consumer optimal trajectories are updated using (\ref{eq:opt-residence}) and copy of the iterate value $\left\{{p}_i^t[l+1]\right\}_{t=1}^{T}$ is sent to the operator. We note that (\ref{eq:opt-residence}) is a MIQP because of the quadratic regularization term ensuring consensus with the operator objective and binary constraints for the EV charging unit. We note that the {\sc REVS} problem which was originally an MILP is converted to a QP for the operator and MIQPs for individual residences using the proposed ADMM based framework.
Once the updated local iterates are exchanged, the operator and residential updates the local dual variables using (\ref{eq:gamma}).

The centralized approach to solve the MILP guarantees convergence to the global optimum solution. However, the concern of sharing private consumer information with the network operator hinders the approach. The proposed ADMM based distributed framework avoids sharing of private and proprietary information and only uses exchange of power consumption data. The approach converts the problem into a QP for the operator and MIQPs for each residence. However, the size of each problem is significantly smaller than the original MILP. The convergence of the algorithm to the optimal solution of (\ref{eq:total-opt}) is formally stated next.

%The iterative procedure is ended when the optimal power trajectory $p_i^t$ computed by residence matches the trajectory $\tilde{p}_i^t$ evaluated by the network operator for all residences $i\in\mathscr{H}$. In such a case, there is no further update in the dual variable. Here $\kappa$ is a suitable value to aid convergence in ADMM techniques as suggested in \cite{sairaj2014

\begin{proposition}
The iterates $\left\{\tilde{\mathbf{p}}^t[l]\right\}_{t=1}^{T}$ and $\left\{\mathbf{p}^t[l]\right\}_{t=1}^{T}$ produced by $[\mathbf{S1}]-[\mathbf{S2}]$ are convergent, for any $\kappa>0$. Further, $$\lim_{l\rightarrow \infty}\left\{\tilde{\mathbf{p}}^t[l]\right\}_{t=1}^{T} = \lim_{l\rightarrow \infty}\left\{\mathbf{p}^t[l]\right\}_{t=1}^{T} = \left\{\mathbf{p}^t_{\textrm{opt}}\right\}_{t=1}^{T}$$
where $\mathbf{p}^t_{\textrm{opt}}$ denotes the optimal power usage trajectory.
\end{proposition}
ADMM has been proved to converge to the optimal solution for convex problems~\cite{boyd}. However, in our paper the problem involves binary constraints for the EV charging unit which renders the problem to be non-convex. Interestingly, ADMM can converge to optimality in the exact sense for non-convex problems with binary variables. Therefore, we can guarantee that the proposed framework converges to the optimal solution.

\section{Experiments}

The experiments are conducted in order to study the effects of EV adoptions at different levels (30\%, 60\%, 90\%). 
We compare effects of two optimization scenarios (individual vs. distributed) on EV scheduling behavior in different communities. 
Under the \emph{individual optimization} scenario, customers charge their EVs based on individual preferences without thinking about the impact on the network. The optimal schedule is obtained by solving (\ref{eq:ind-opt}) for each EV adopter.
With \emph{distributed optimization}, the customers agree to an optimal EV charging schedule where they are benefited as well as the network reliability is maintained. This allows the network operator to maintain acceptable node voltage level throughout the network. The optimal schedule is obtained by iteratively solving Equations (\ref{eq:opt-operator}), (\ref{eq:opt-residence}), and (\ref{eq:gamma}) until convergence.
  
% Particularly, we aim to identify how much EV adoption is acceptable using the existing distribution network infrastructure.  
% An important aspect of network reliability is maintaining line flows and  node voltages within an acceptable range : ($0.95-1.05$ p.u)~\cite{ansi}. 
%Node voltages outside this range is considered to be a \emph{reliability} concern for the network operator.
%\swapna{Please add description of edge flow/line rating and its importance to maintaining network reliability}
Particularly, we aim to compare the reliability of the network when these two methods are used to schedule residential EV charging for varied levels of EV adoption. Note that network reliability is the ability to operate with edge power flows within the line capacities and node voltage within the bandwidth ($0.95-1.05$ p.u)~\cite{ansi}. Hence, these two measures -- node voltage and edge power flow are used to quantify the impact of network reliability at different levels (30\%, 60\%, 90\%) of EV adoptions in multiple communities in the distribution network.
% Two experiments are conducted on the distribution network.

% \noindent
% \emph{Experiment 1.} Comparison of edge flows at different \% of adoptions in different communities under distributed and individual optimization scenarios. 

% \noindent
% \emph{Experiment 2.} Comparison of node voltages (voltage at residence level) at different \% of adoptions in different communities under distributed and individual optimization scenarios.

\begin{table}[htbp]
\centering
\caption{Hourly electricity rate for the experiments}
\label{tab:price-summer}
\begin{tabular}{ccccc}
\hline
\textbf{\begin{tabular}[c]{@{}c@{}}Time \\ interval \\ (HH:MM)\end{tabular}} & \begin{tabular}[c]{@{}c@{}}00:00-\\ -05:00\end{tabular} & \begin{tabular}[c]{@{}c@{}}05:00-\\ -15:00\end{tabular} & \begin{tabular}[c]{@{}c@{}}15:00-\\ -18:00\end{tabular} & \begin{tabular}[c]{@{}c@{}}18:00-\\ -00:00\end{tabular} \\ \hline
\textbf{\begin{tabular}[c]{@{}c@{}}Cost \\ (\$/kWhr)\end{tabular}} & 0.07866 & 0.09511 & 0.21436 & 0.09511 \\ \hline
\end{tabular}
\end{table}

\begin{figure*}[!ht]
    \centering
    \includegraphics[width=0.4\textwidth]{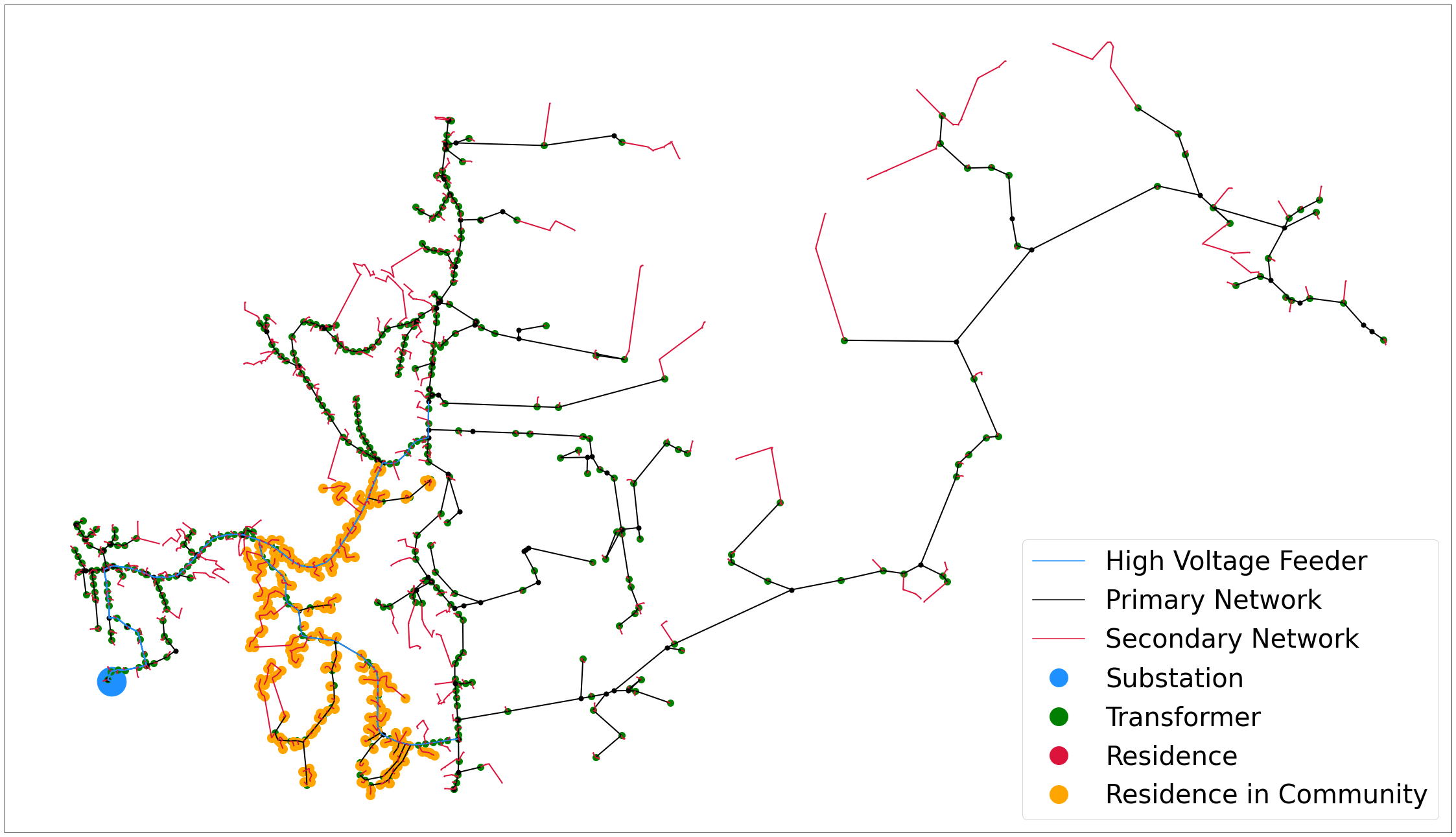}
    \includegraphics[width=0.58\textwidth]{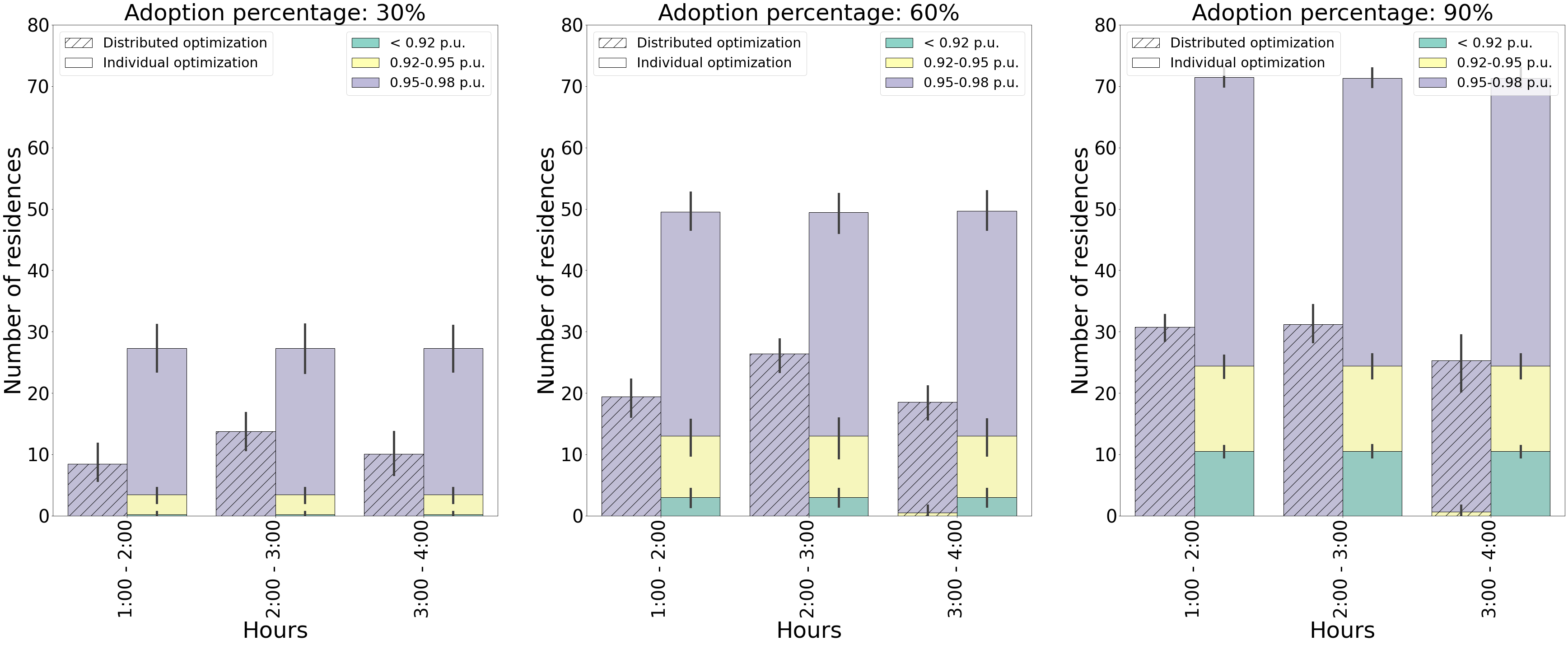}
    \includegraphics[width=0.4\textwidth]{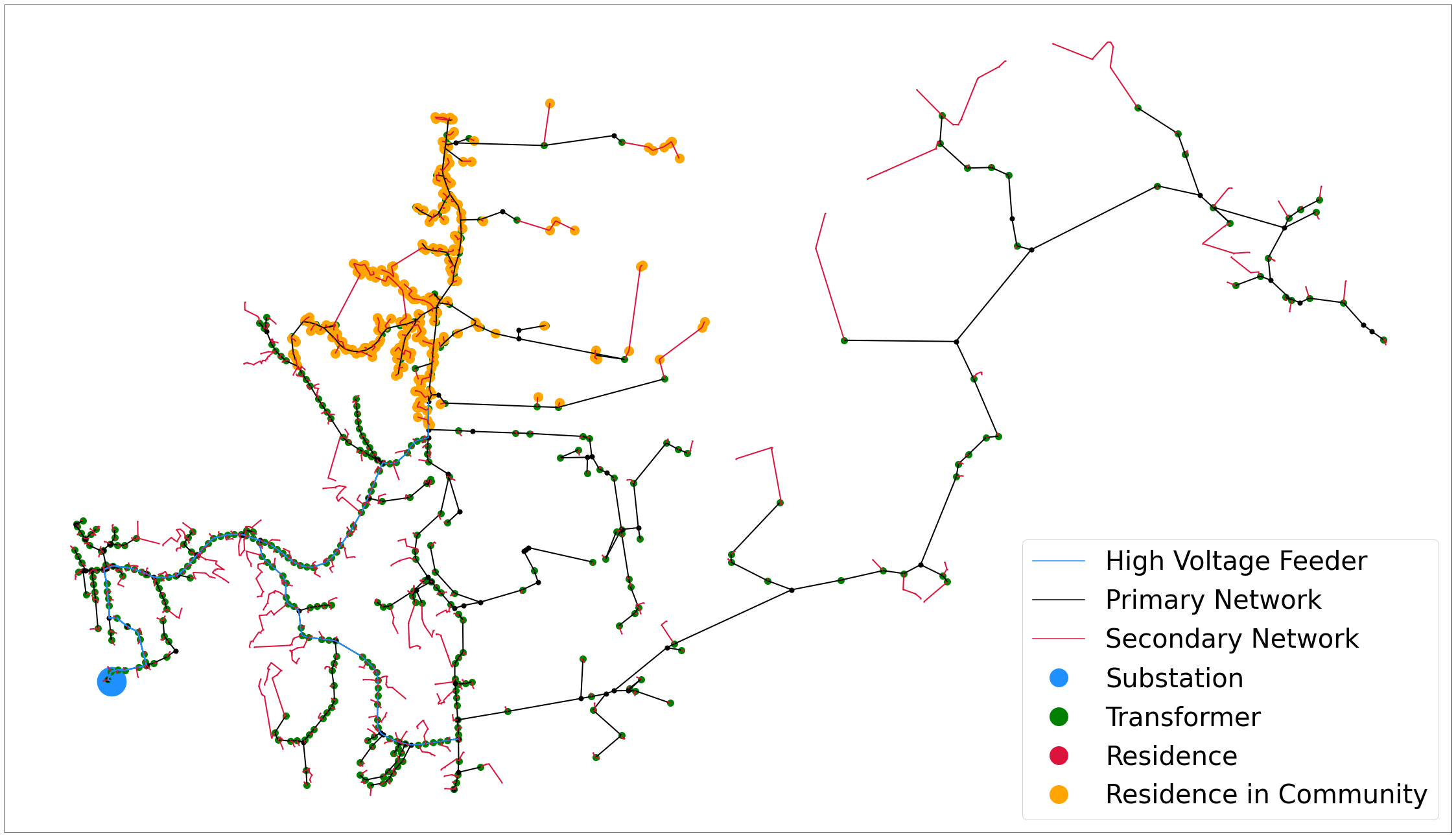}
    \includegraphics[width=0.58\textwidth]{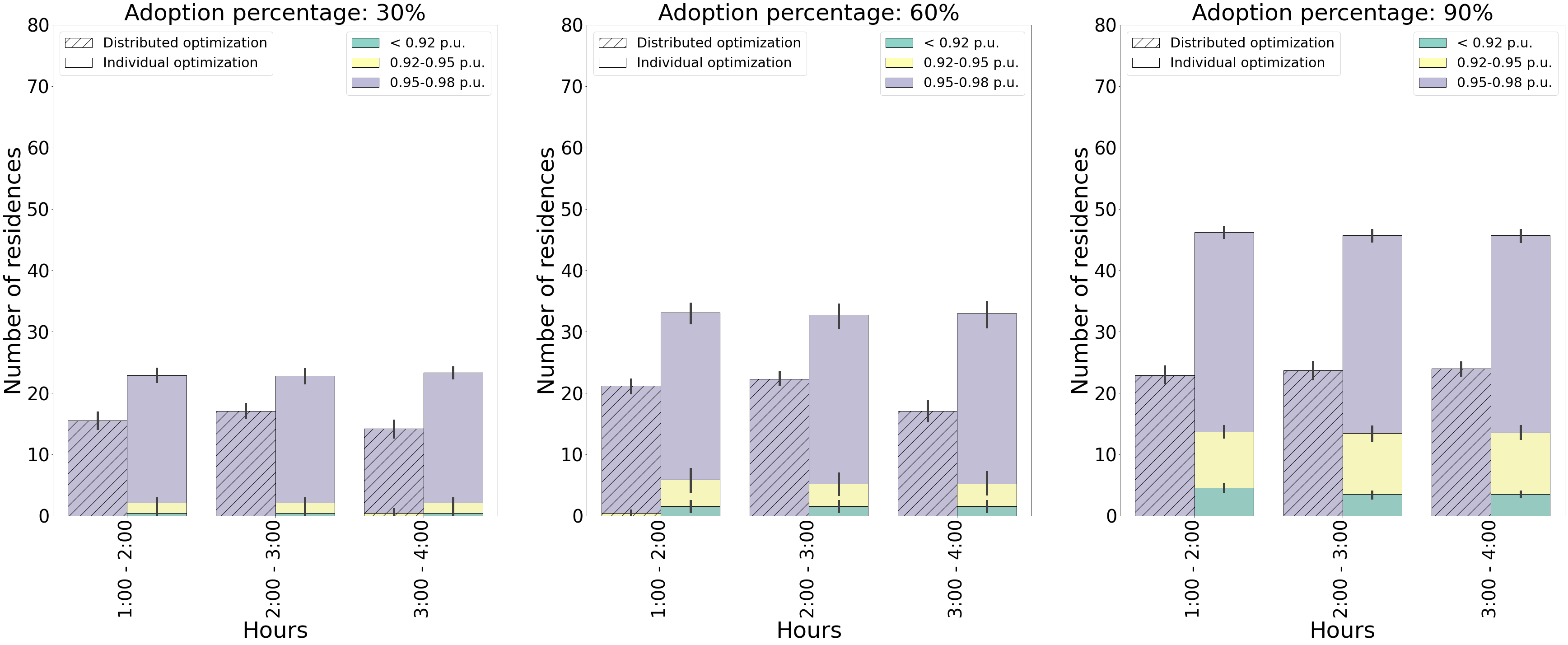}
    \caption{Impact of residential EV charging is analyzed for two different residential communities within the same network: `Com-A'(top) and `Com-B'(bottom). The orange nodes shown in the network denote the residences in the two communities. The individual optimization leads to undervoltage (less than 0.95 p.u.) at a significant number of residences. This can be avoided by using the proposed distributed optimization method even for higher levels of EV adoption.}
    \label{fig:adoption-out-limit}
\end{figure*}

A small area of Montgomery county in Virginia is considered as the region of interest for our study. 
Household level synthetic hourly electricity consumption profiles are used. These timeseries are created using several population surveys, statistical models, and physics based models of household devices and validated using real data~\cite{swapna2018}. The data also has household level demographics and spatial attributes.
Synthetically generated distribution networks created using electrical engineering concepts and resembling actual networks are used for the purpose of our analysis~\cite{rounak2020}. 
Hourly electricity rate (in \$/kWhr) is known to the residential customers (Table~\ref{tab:price-summer}). 
%Experiments are conducted for a typical summer day. 
Time of use (TOU) hourly electricity rate provided by the off-peak plan of a utility company serving the particular geographical region~\cite{dominion_tou} are used. 

%\swapna{If space permits, we might need to add references to these assumed values.}
\emph{Assumptions.}~All EVs have a uniform charge capacity of $20$kWhr and are available to be charged between 4:00p.m. and 5:00a.m. The initial state of charge is assumed to be $20\%$ and the EVs are required to be charged to at least $90\%$. 
A uniform rating of $4.8kW$ of the residential EV charger. 
%The charge capacity is chosen based on the latest EV models which can be used for regular commuting purposes within $100$km. 
Households are randomly selected as EV adopters in the network.
All adopters have necessary provisions to charge their EVs at their residential premises.

\section{Results}
\begin{figure}[!ht]
    \centering
    \includegraphics[width=0.48\textwidth]{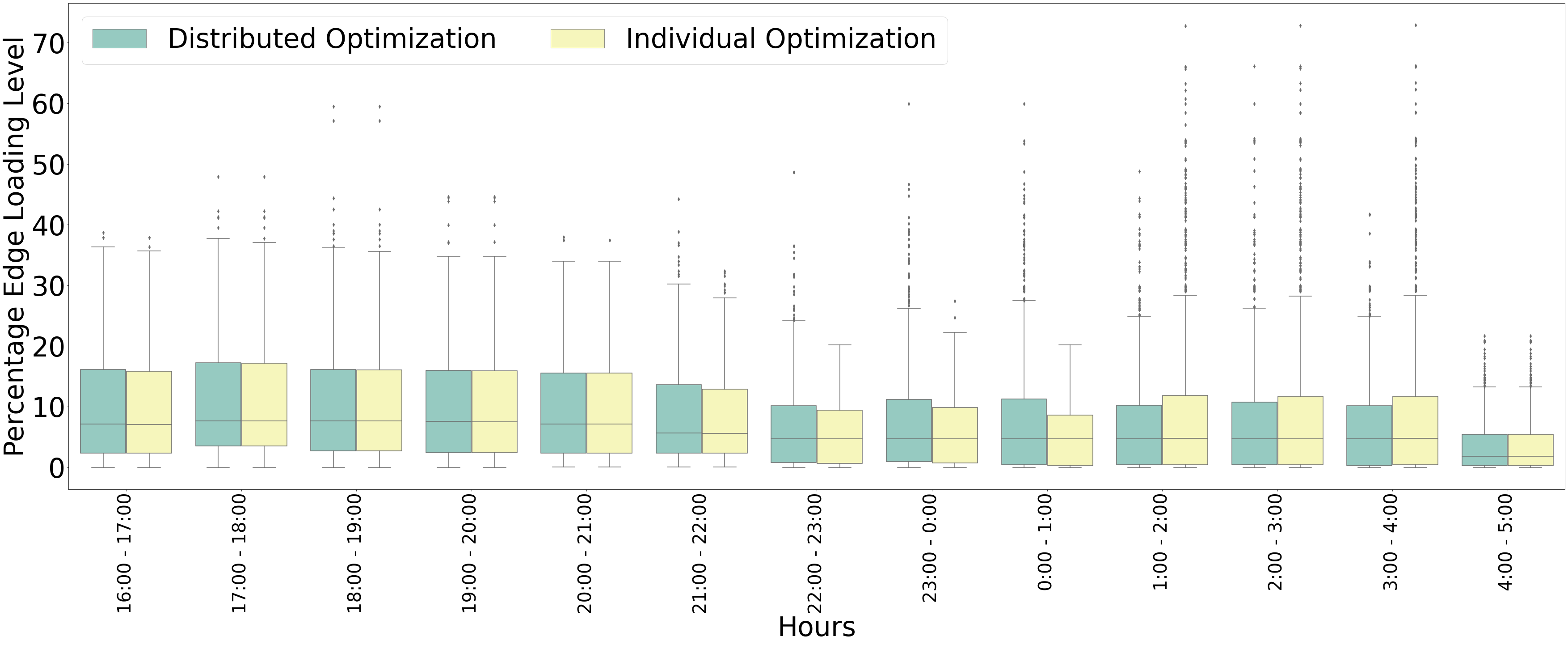}
    \includegraphics[width=0.48\textwidth]{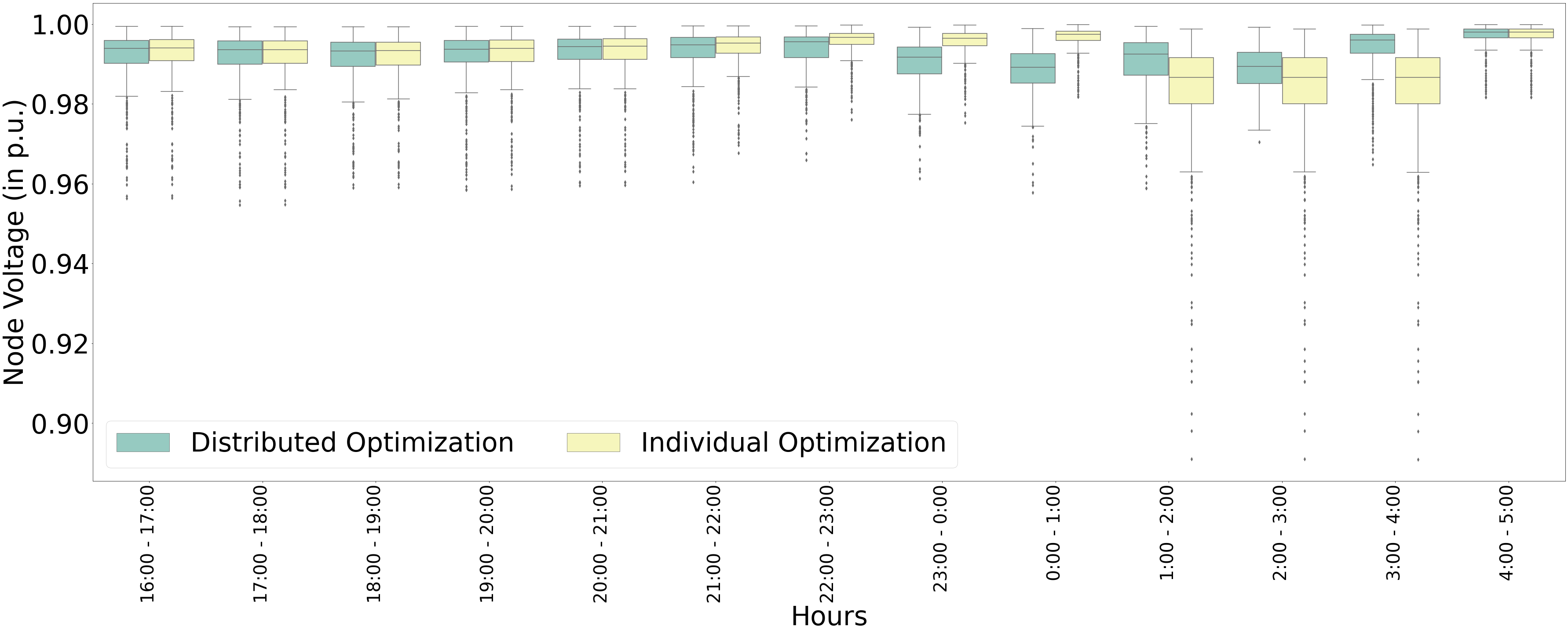}
    \caption{Comparison of line loading level (edge flows) and node voltages for residential EV adoption of $90\%$ in `Com-A' of network. The high EV adoption does not significantly affect the line loading level. However, node voltages at multiple residences in the network are outside the acceptable voltage limits of $0.95-1.05$ p.u.}
    \label{fig:flow-vs-volt}
\end{figure}

Figure~\ref{fig:flow-vs-volt} describes edge power flow as a percentage of line capacities and node voltages in `Com-A' community in the network when EV adoption has reached 90\%.
Figure~\ref{fig:flow-vs-volt} (top) shows that the edge flow (line rating capacity) levels in the network are well within limits even when 90\% of residences have adopted EV. 
However, the same cannot be observed for node voltages in the network.
Figure~\ref{fig:flow-vs-volt} (bottom) shows that the node voltages at several residences are outside the acceptable limits.
We notice that maximum number of node voltage violations occur in the period where the hourly electricity rate (in \$/kWhr) is minimum.

We further explore effects of node voltage violation, at different adoption levels in two communities (`Com-A', `Com-B') of residences in the network.
Figure~\ref{fig:adoption-out-limit} shows the selected communities in the network and node voltage violation at different adoption levels.
The results are obtained after performing optimizations under two scenarios : individual and distributed.
We focus our attention only on the time intervals where the hourly electricity rate (in \$/kWhr) is minimum (i.e. time windows of maximum node violations). 
The node voltage violation is divided into 3 ranges : less than $0.92$ p.u., between $0.92-0.95$ p.u. and between $0.95-0.98$ p.u. 
Though the last voltage range is not considered as voltage violation by the practised ANSI standard~\cite{ansi}, it can be considered as a sign of reduced network reliability.
The clustered bar chart shows significantly higher  number of residences with voltage violation under individual optimization scenario as compared to the distributed optimization scenario.
The number of residences with voltage less than $0.95$ p.u. is close to zero at all considered time intervals. 
The observations are similar for both communities in the network.
This shows that, if the proposed distributed framework is used to schedule EV charging units, the network operator is able to dispatch the power without compromising on system reliability.

% Various communities
Figure~\ref{fig:adoption-out-limit} shows that under the individual optimization approach the number of residences with undervoltage during the cheap electricity hours increases with an increase in the level of adoption. However, this trend is not consistent in the distributed optimization approach. This is because the later approach also ensures system reliability along with consuming electricity during cheap hourly rates. 
The distributed framework does this by allocating small amount of EV charging during time intervals where the hourly electricity rates are relatively higher.

We also notice that the number of residences experiencing undervoltage issues for the same level of adoption differs significantly when we consider different communities for EV adoption. These differences can be attributed to location and energy usage of adopter households in the network and the resulting voltages at different nodes.
The error bars on the bar chart (Figure~\ref{fig:adoption-out-limit}) show variation in number of residences violating node voltage in each category after multiple runs of the adoption level.

\section{Conclusion}
In this paper we propose an ADMM method based distributed framework for scheduling EV charging for consumers at residential premises while maintaining reliability of the network. 
We observe that individual consumer loads, spatial layout of the network, and electricity pricing all affect network reliability.
We provide a set of results on a small distribution network in Montgomery county of Virginia, USA for EV adoption at different levels in multiple residential communities. 
We inspect the network reliability with two measures : node voltage and edge flows (line rating/line loading).
Results show that the proposed distributed approach for scheduling residential EV charging enables the network operator to maintain system reliability by respecting line limits and having fewer undervoltage households and almost no households with node violation even for higher levels of EV adoption as opposed to an individual approach.

%However, this work does not include detailed EV usage models where the adopters use the EVs for commuting throughout the day. Such models need to include activity patterns of users and would be pursued for a future direction of research. Furthermore, the spatial dependency between network reliability and adopting communities has been observed from our results. This poses us a problem of identifying the critical locations of EV adopters which maximizes the loss of system reliability and thereby help system planners to strengthen the distribution network infrastructure accordingly.
% \begin{enumerate}
%     \item propose a distributed framework for scheduling EV load in the  power distribution network. The need for a framework for load scheduling arises in the distribution network as the grid undergoes changes and larger adoption of EV
%     \item Simulate  scenarios that DO WHAT. The results suggest that --
%     \item limitations: more accurate models for EV charging where adopters use the EVs during charging hours.
% \end{enumerate}

\bibliographystyle{named}
\bibliography{ijcai22}

\clearpage
\section*{Appendix}

\subsection*{Preliminaries of power networks}
\begin{figure}[htbp]
    \centering
    \includegraphics[width=0.235\textwidth]{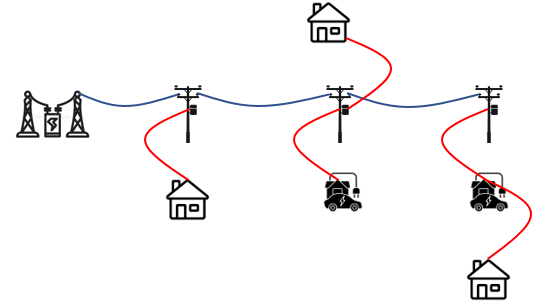}
    \includegraphics[width=0.235\textwidth]{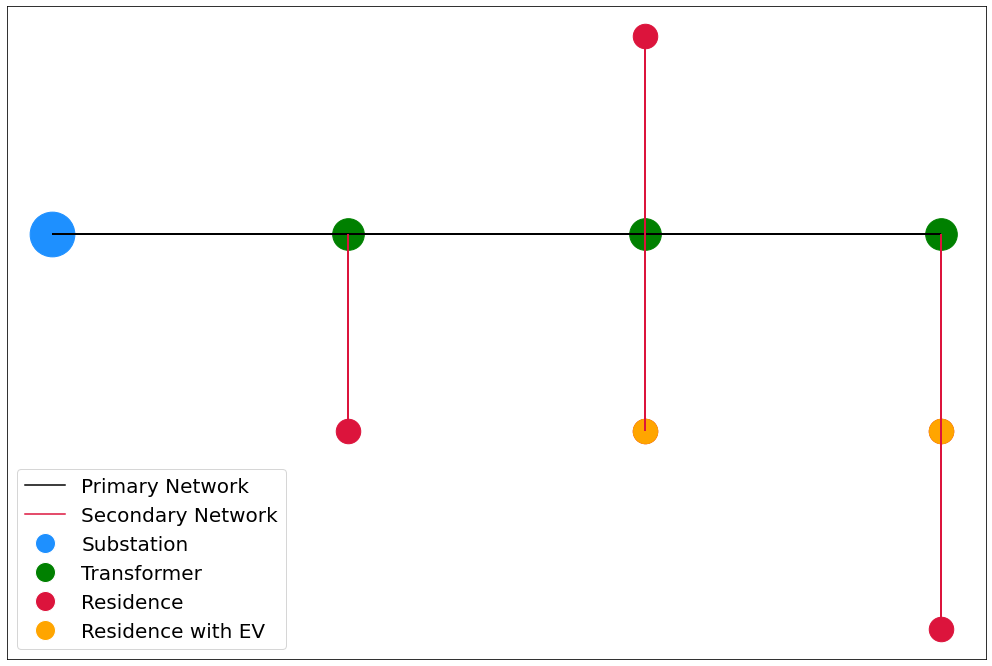}
    \caption{Schematic of power distribution network with few residential EV adopters.}
    \label{fig:intro}
\end{figure}

Fig.~\ref{fig:intro} shows a typical power distribution network with few EV adopters. The network is a tree rooted at the substation node and connects residences through local pole top transformers. 
A branch originating from the substation is also termed as a `feeder'. 
The power flowing along each edge results in voltage drop between the nodes. 
Therefore, the node at the extreme end of the feeder experiences the minimum voltage. 

\subsection*{Convergence of ADMM methodology}
In the paper, we have discussed the need of a distributed methodology to respect the privacy of consumer information. However, this alters the original MILP problem into one QP and multiple MIQPs. Here, we discuss the deviation of the optimal solution obtained by the distributed framework from that obtained by a centralized approach. Note that the centralized approach involves solving the MILP in (\ref{eq:total-opt}).
\begin{figure}[htbp]
    \centering
    \includegraphics[width=0.48\textwidth]{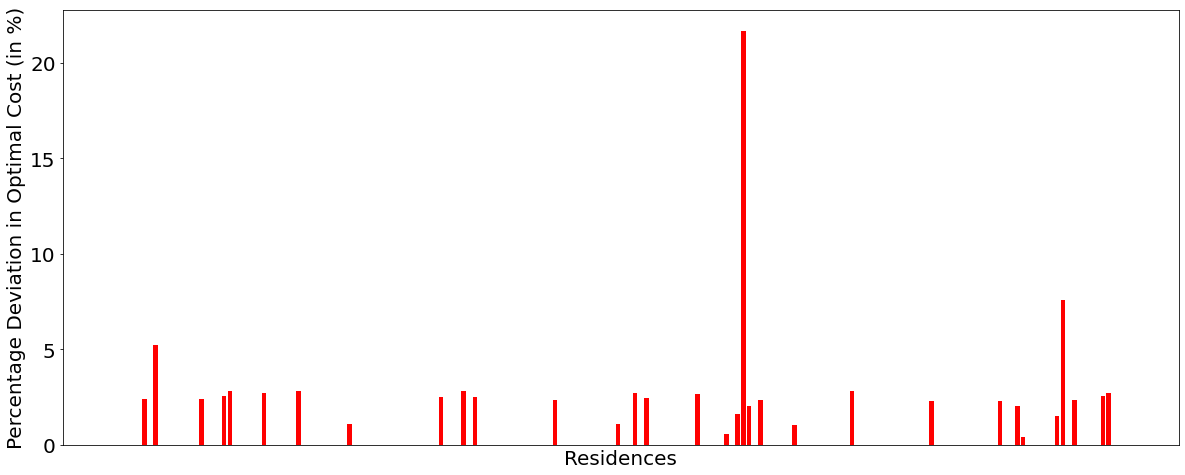}
    \caption{Percentage deviation of the optimal solution obtained using the proposed framework from the solution obtained by centrally solving (\ref{eq:total-opt})}
    \label{fig:dev}
\end{figure}
Fig.~\ref{fig:dev} shows the percentage deviation in the optimal cost incurred by the EV adopting residences. This is done for an adoption level of $50\%$ in `Com-A' of the network. The solution obtained by the proposed distributed approach converges with the solution obtained by the centralized approach for majority of residences. For the other residences the deviation in the optimal cost lies mostly below $5\%$ with one residence having a large deviation of $20\%$. Therefore, we can conclude that the optimal solution converges for most of the residences. The deviation is the price we pay to respect the privacy of consumers while dealing with this problem.

\subsection*{Additional Results}
\begin{figure}[htbp]
    \centering
    \includegraphics[width=0.48\textwidth]{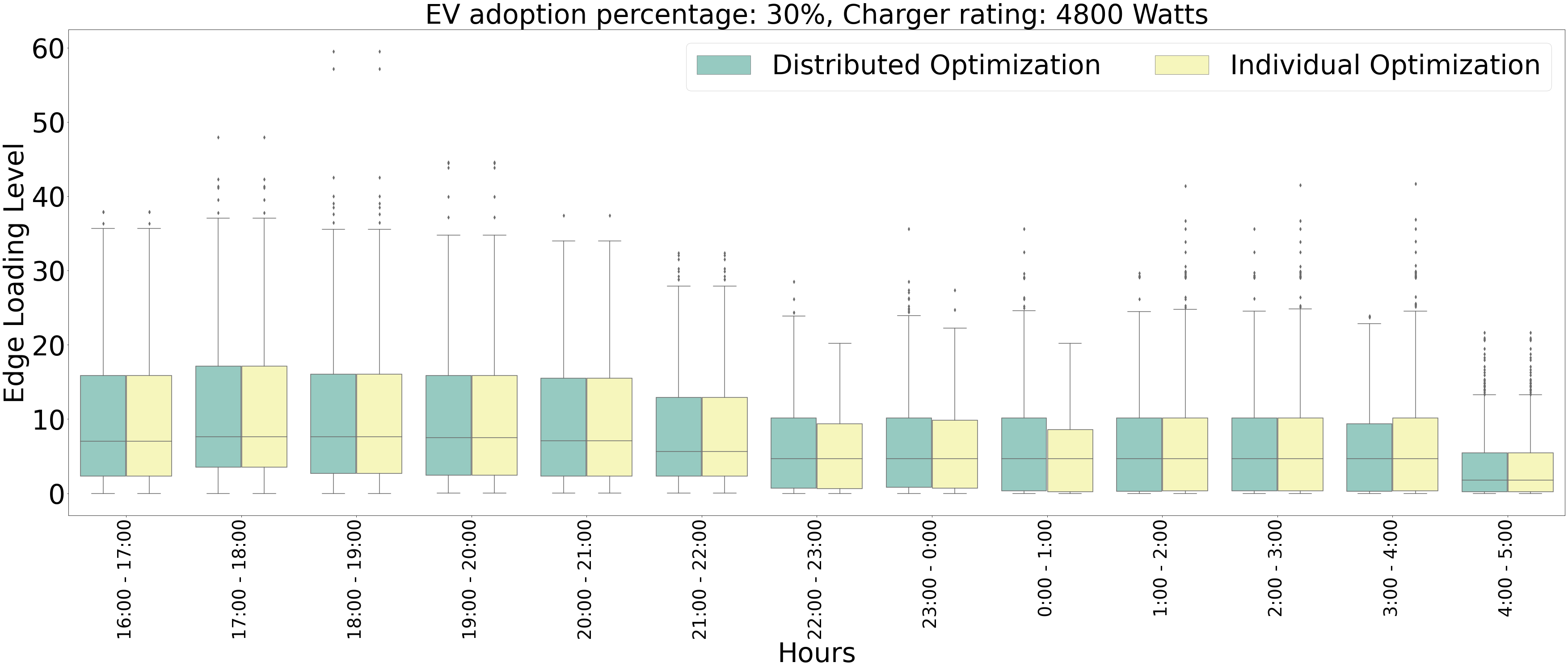}
    \includegraphics[width=0.48\textwidth]{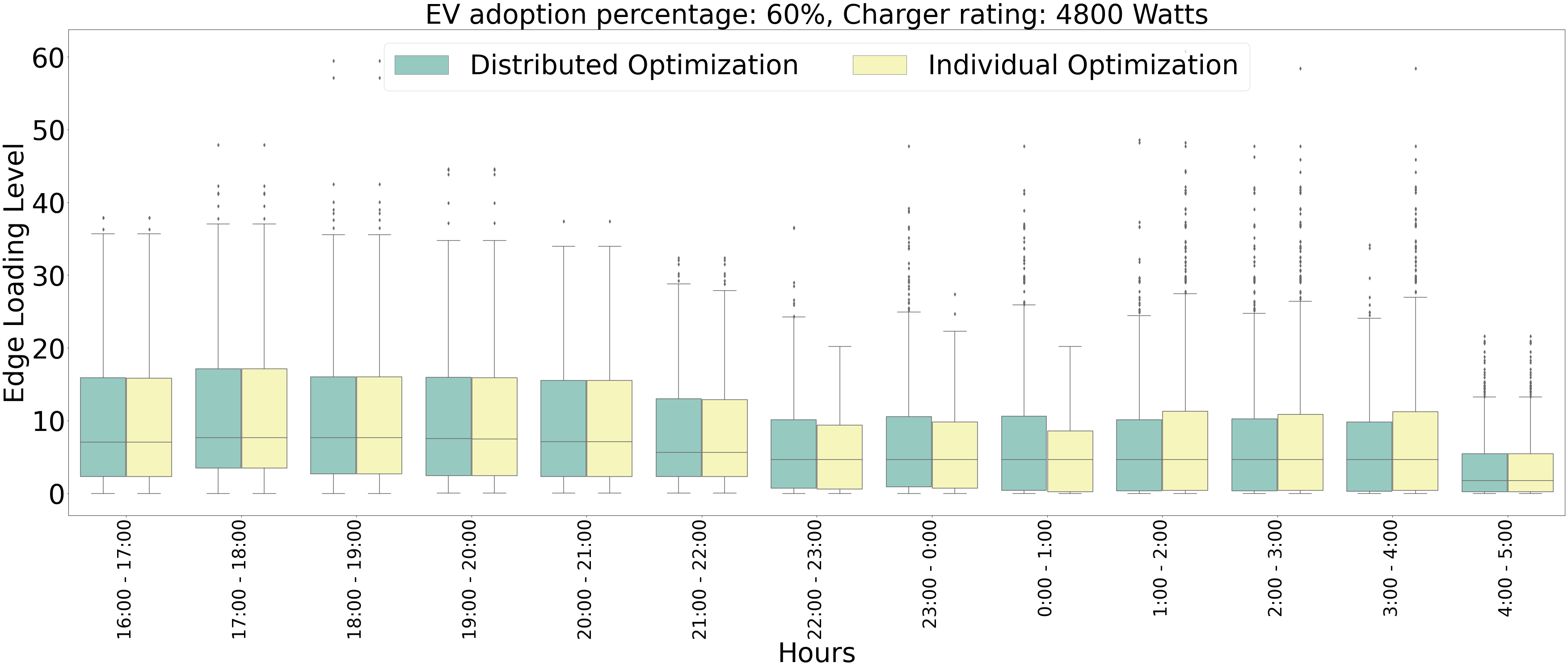}
    \caption{Comparison of line loading level (edge flows) for residential EV adoption of $30\%$ and $60\%$ in `Com-A' of network. The EV adoption does not significantly affect the line loading level for either of the optimization methods.}
    \label{fig:extra-1}
\end{figure}

\begin{figure}[htbp]
    \centering
    \includegraphics[width=0.48\textwidth]{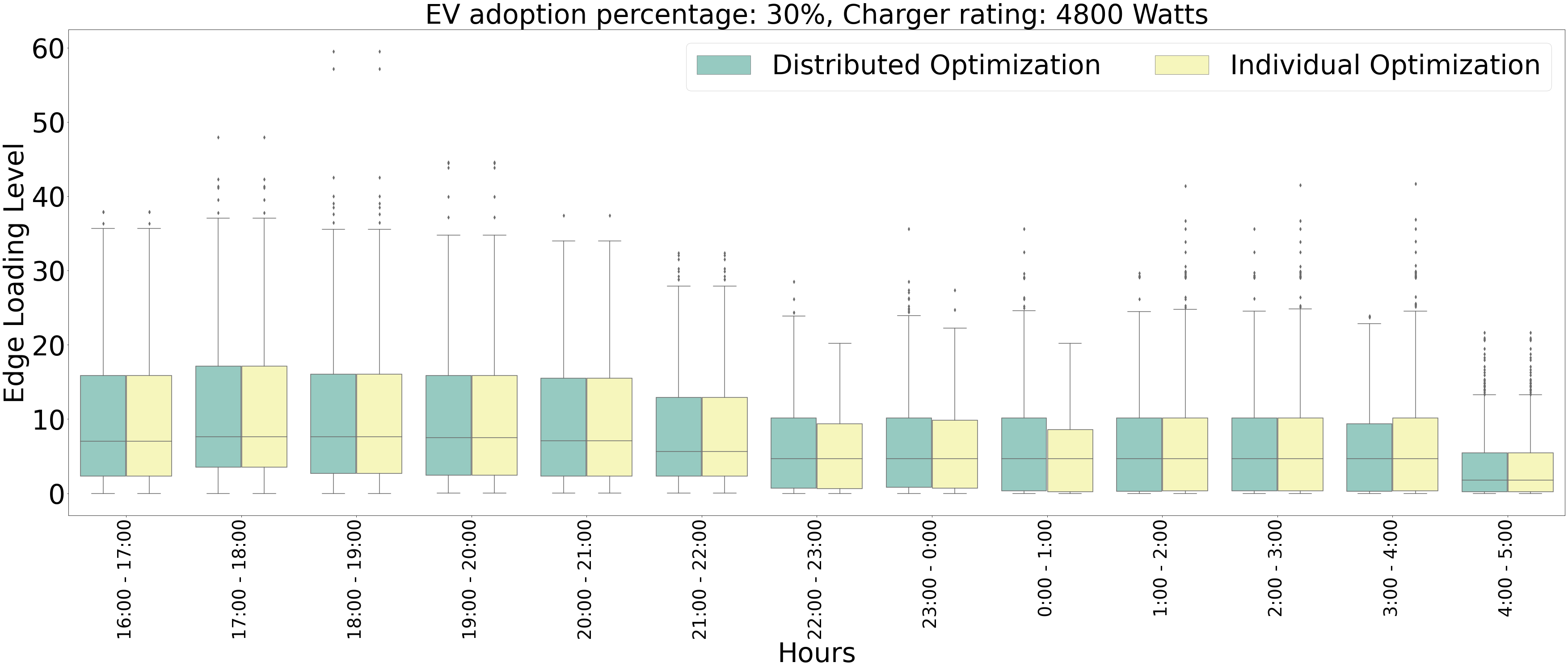}
    \includegraphics[width=0.48\textwidth]{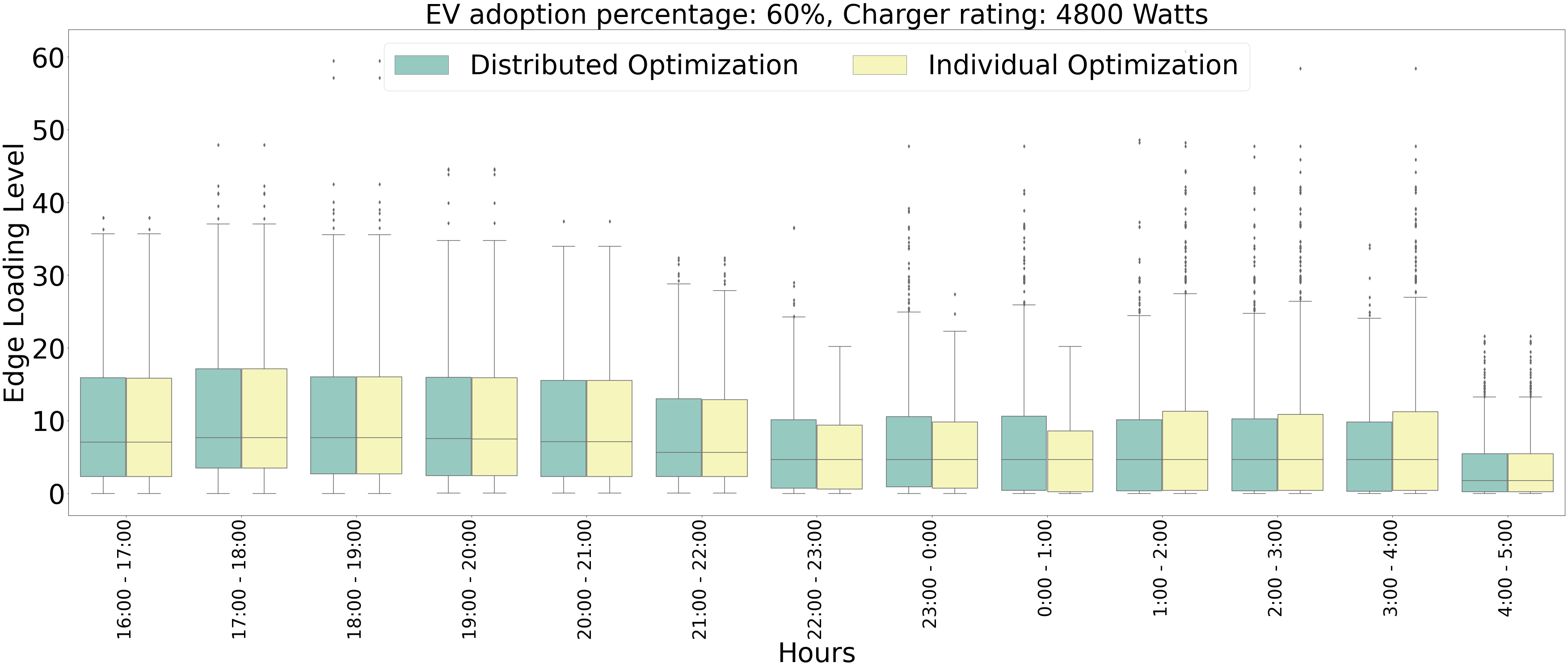}
    \caption{Comparison of residence node voltages for residential EV adoption of $30\%$ and $60\%$ in `Com-A' of network. The EV adoption adversely affects the node voltages during cheap electricity hours for the individual optimization scenario.}
    \label{fig:extra-2}
\end{figure}

We provide the plots for line loading levels and node voltages for multiple adoption levels in the two communities Fig.~\ref{fig:extra-1} and Fig.~\ref{fig:extra-2}. In all the cases, we note that the edges in the network are not loaded to their full capacities. However, the voltages of multiple residences fall outside the acceptable limits of $0.95-1.05$p.u.

\begin{figure*}[htbp]
    \centering
    \includegraphics[width=0.4\textwidth]{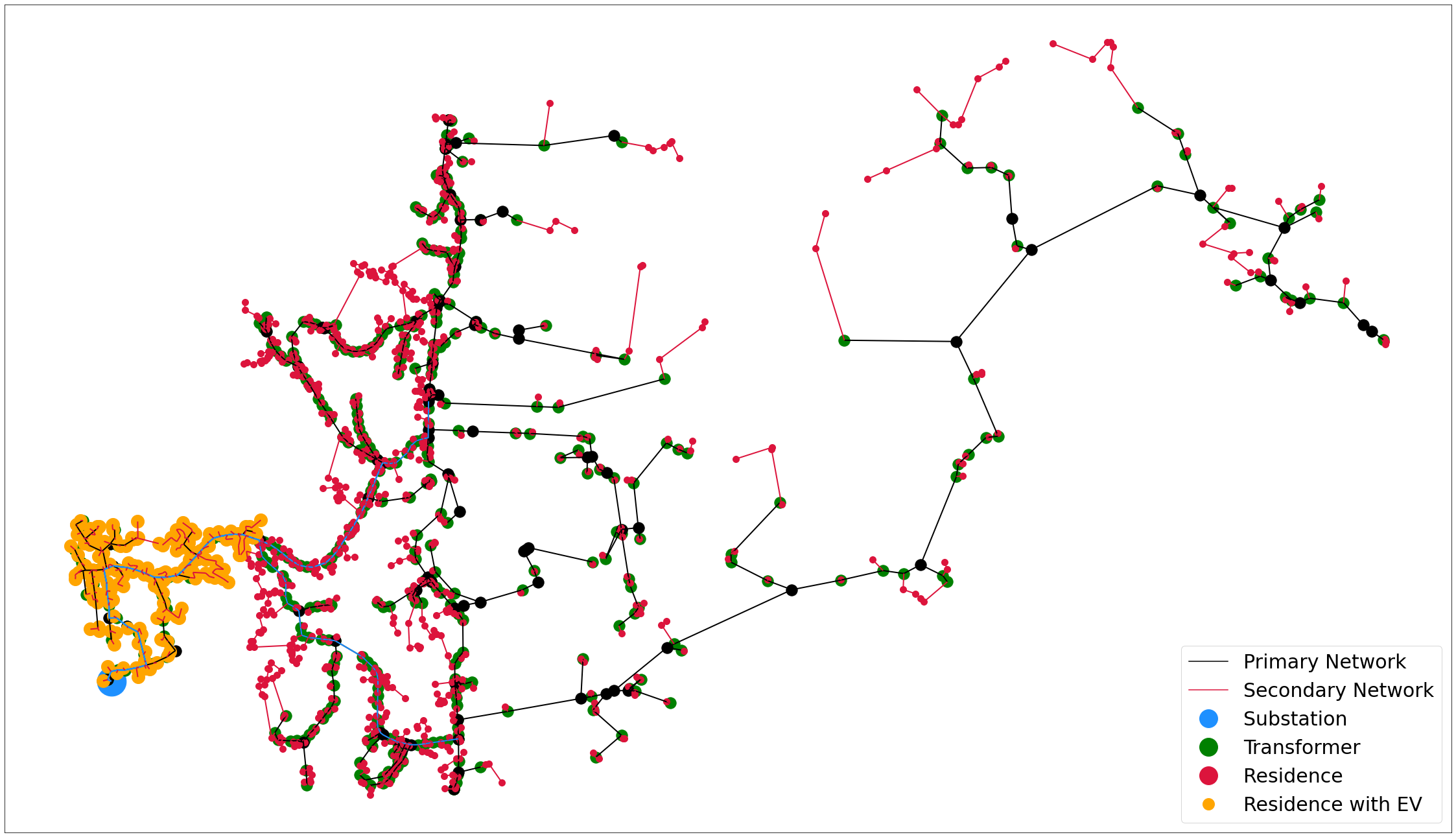}
    \includegraphics[width=0.58\textwidth]{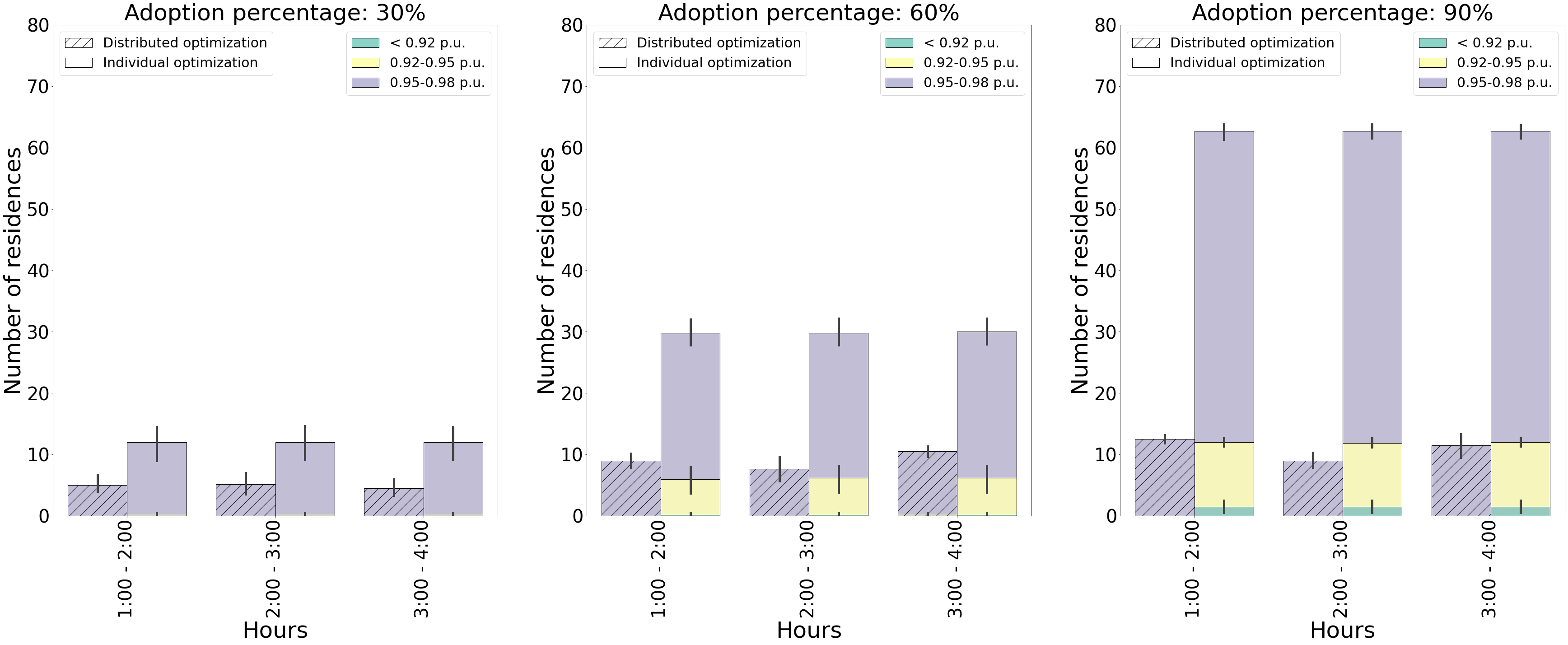}
    \includegraphics[width=0.4\textwidth]{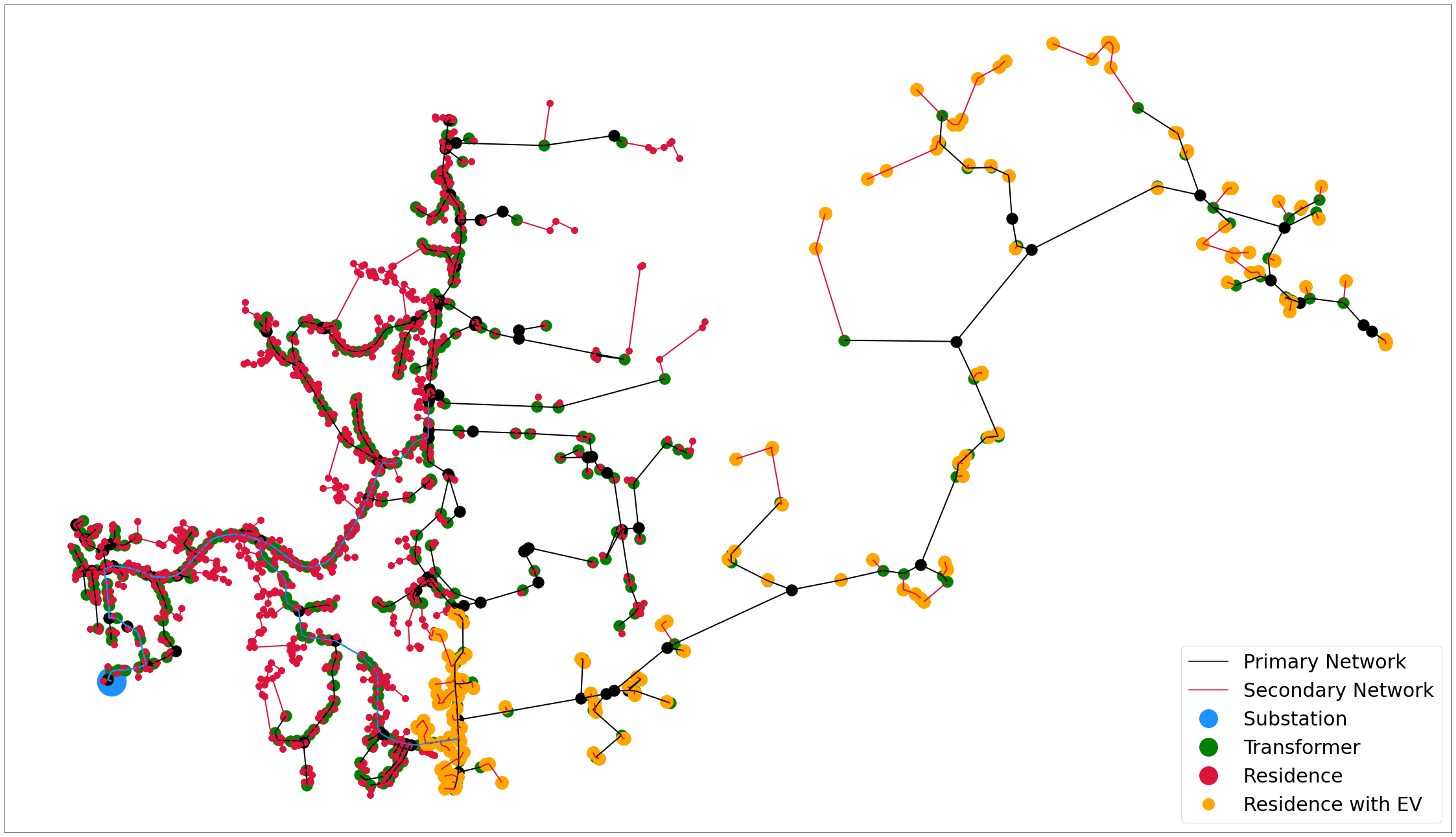}
    \includegraphics[width=0.58\textwidth]{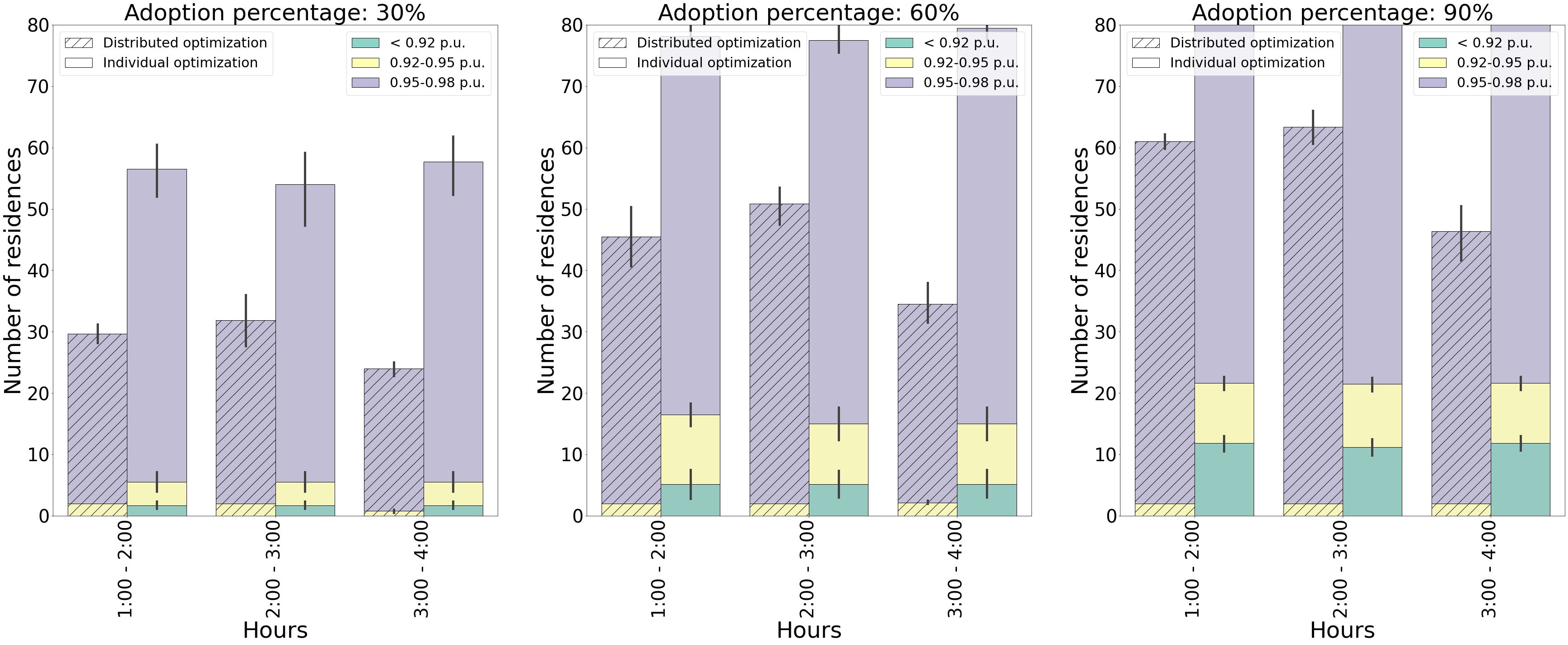}
    \includegraphics[width=0.4\textwidth]{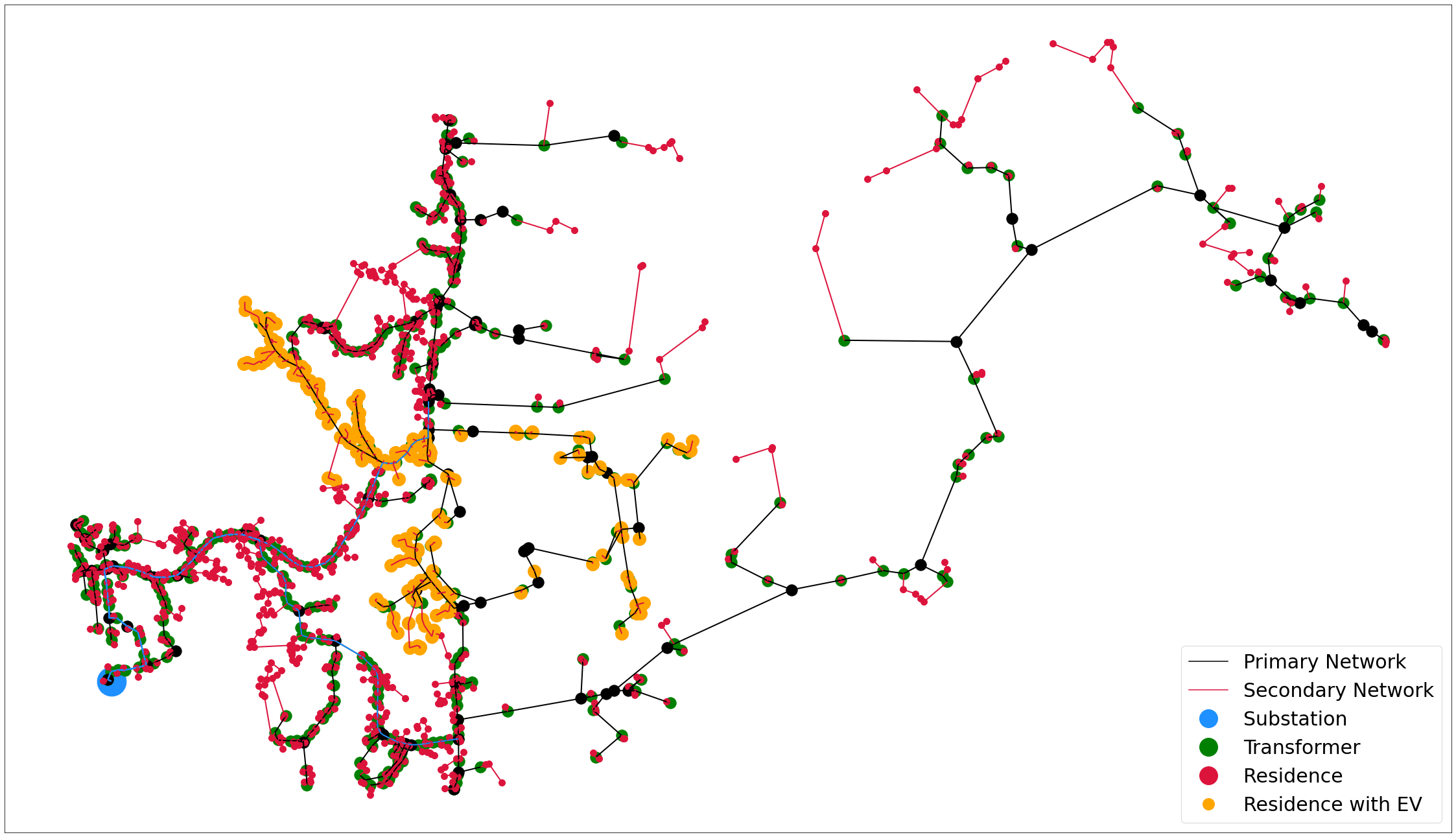}
    \includegraphics[width=0.58\textwidth]{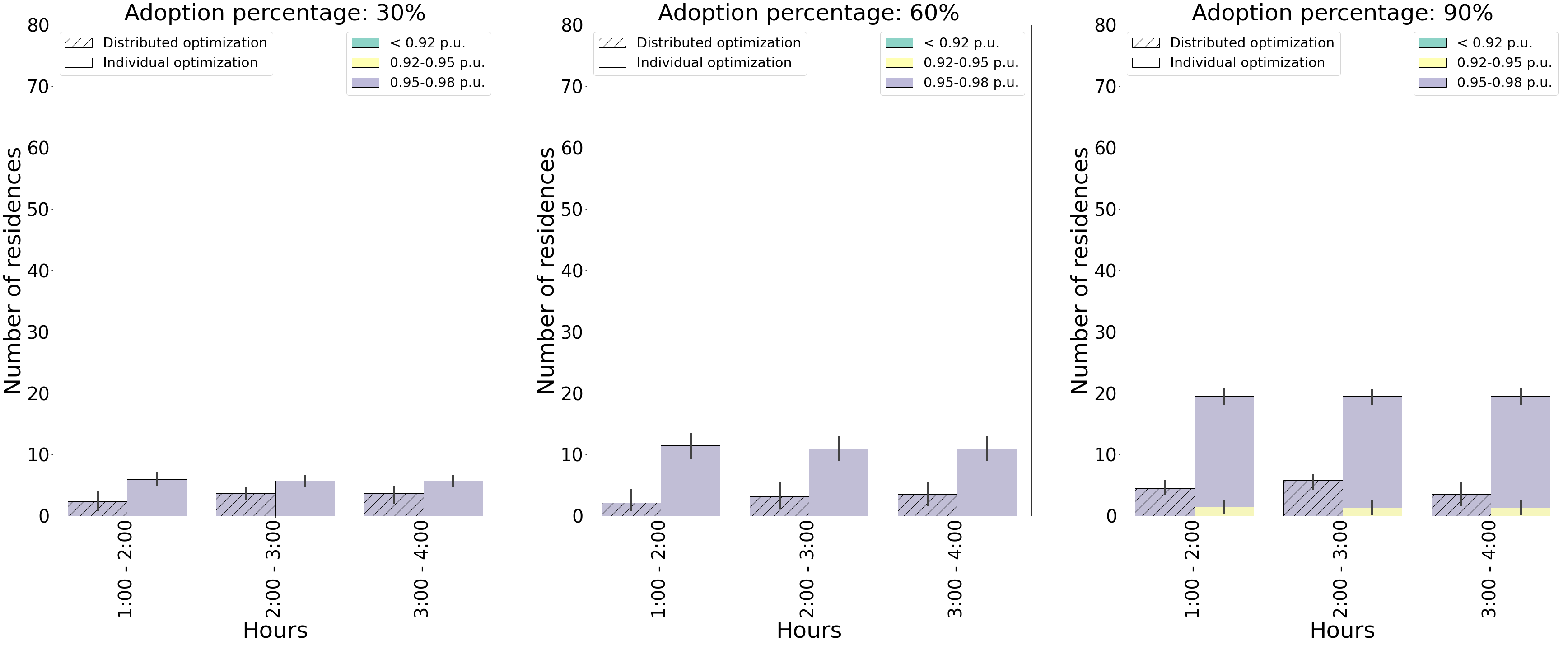}
    \caption{Impact of residential EV charging is analyzed for three more residential communities within the same network: `Com-C'(top), `Com-D'(middle) and `Com-E'(bottom). The orange nodes shown in the network denote the residences in the two communities. The individual optimization leads to undervoltage (less than 0.95 p.u.) at a significant number of residences. This can be avoided by using the proposed distributed optimization method even for higher levels of EV adoption.}
    \label{fig:extra-3}
\end{figure*}
We further compare the distributed and individual optimization frameworks for three more residential communities in the network. These are shown in Fig.~\ref{fig:adoption-out-limit}. We note that the impact on reliability of the network in terms of number of residences experience undervoltage is different for each community in the network. This proves the spatial dependency of network reliability on EV adoption location.
\end{document}